\documentclass[%
 aip,
 amsmath,amssymb,
 reprint,%
 nofootinbib,
]{revtex4-1}

\usepackage{graphicx}% Include figure files
\usepackage{dcolumn}% Align table columns on decimal point
\usepackage{bm}% bold math
\usepackage[utf8]{inputenc}
\usepackage[T1]{fontenc}
\usepackage{mathptmx}
\usepackage{textcomp}
\usepackage{gensymb}
\usepackage{placeins}
\usepackage{appendix}
\usepackage{mdwlist}

\begin{document}

\preprint{APS/123-QED}

\title{Thermal Kinetic Inductance Detectors for Millimeter-Wave Detection}
\author{A. Wandui}
 \email{awandui@caltech.edu.}
\affiliation{ Department of Physics, California Institute of Technology, Pasadena, CA, 91125, USA}
\author{J.J. Bock}
 \altaffiliation[Also at ]{Jet Propulsion Lab, Pasadena, CA, 91109, USA}
\affiliation{ Department of Physics, California Institute of Technology, Pasadena, CA, 91125, USA}
\author{C. Frez}
\affiliation{Jet Propulsion Lab, Pasadena, CA, 91109, USA}
\author{M. Hollister}
\affiliation{Fermilab, Batavia, IL, 60510, USA}
\author{L. Minutolo}
\affiliation{ Department of Physics, California Institute of Technology, Pasadena, CA, 91125, USA}
\author{H. Nguyen}
\affiliation{Jet Propulsion Lab, Pasadena, CA, 91109, USA}
\author{B. Steinbach}
\affiliation{ Department of Physics, California Institute of Technology, Pasadena, CA, 91125, USA}
\author{A. Turner}
\affiliation{Jet Propulsion Lab, Pasadena, CA, 91109, USA}
\author{J. Zmuidzinas}
\affiliation{ Department of Physics, California Institute of Technology, Pasadena, CA, 91125, USA}
\author{R. O'Brient}
  \altaffiliation[Also at ]{ Department of Physics, California Institute of Technology, Pasadena, CA, 91125, USA}
\affiliation{Jet Propulsion Lab, Pasadena, CA, 91109, USA}

\date{\today}% It is always \today, today,
             %  but any date may be explicitly specified

\begin{abstract}
Thermal Kinetic Inductance Detectors (TKIDs) combine the excellent noise performance of traditional bolometers with a radio frequency (RF) multiplexing architecture that enables the large detector counts needed for the next generation of millimeter-wave instruments. In this paper, we first discuss the expected noise sources in TKIDs and derive the limits where the phonon noise contribution dominates over the other detector noise terms: generation-recombination, amplifier, and two-level system (TLS) noise. Second, we characterize aluminum TKIDs in a dark environment. We present measurements of TKID resonators with quality factors of about $10^5$ at 80 mK. We also discuss the bolometer thermal conductance, heat capacity, and time constants. These were measured by the use of a resistor on the thermal island to excite the bolometers. These dark aluminum TKIDs demonstrate a noise equivalent power NEP = $2 \times 10^{-17} \mathrm{W}/\mathrm{\sqrt{Hz}} $ with a $1/f$ knee at 0.1 Hz, which provides background noise limited performance for ground based telescopes observing at 150 GHz.

\end{abstract}
  
\maketitle

\section{\label{sec:introduction} Introduction}

Cryogenic bolometers and calorimeters have found widespread use in a wide range of scientific applications including dark-matter searches\cite{superCDMS, edelweiss, cresst, cosinus} and neutrino detection\cite{cuore, holmes, echo}, thermal remote imaging and security cameras\cite{Luukanen_2012, May_2013}; and astronomy with wavelengths ranging from millimeter waves to gamma rays\cite{bicep3, spt3g, polarbear, advancedACT, class, piper, spider, ebex2, optical, xraygamma}. Modern bolometric and calorimetric instruments use Transition Edge Sensors (TESes)\cite{irwinhilton2005} for their thermometers because they can be fabricated using thin-film lithography and thus integrated into monolithic arrays with large detector counts.  In practice, in order to limit the thermal cable load on the cryogenic stages, which have limited cooling capacity, multiple TESes are read out on a single line using Superconducting Quantum Interference Devices (SQUIDs)\cite{squids}. SQUIDs are low noise and have low impedance as well as sufficient bandwidth to multiplex many TESes on a single line with minimal crosstalk. The two main schemes developed to multiplex TES pixels are time-division multiplexing (TDM)\cite{timedomainmultiplexing, timedomainmultiplexing2} and frequency-division multiplexing (FDM)\cite{frequencydomainmultiplexing}.  Currently operating instruments fielding kilo-pixel detector arrays have implemented TDM and FDM readout schemes\cite{bicep3, advancedACT, polarbear, spt3g, ebex}.

The next generation of cryogenic instruments under development will require even higher detector counts\cite{biceparray, cmbs4}. This is especially true for applications in which the detectors are designed to be background noise limited. In this case, gains in instrument sensitivity can only be achieved by increasing the number of detectors. However, it is challenging and expensive to integrate and read out arrays with detector counts above $\sim$10,000 using the existing SQUID-based multiplexing schemes.

Kinetic Inductance Detectors (KIDs)\cite{Day2003, Doyle2008, MazinARCONS2013, McKenneyMAKO2012, PatelMicroSpec2013, Golwala_2012, MonfardiniNIKA2011, DoberBLASTTNG2014, KovacsSuperSpec2012} are an alternative detector technology that offer a promising solution to the problem of scaling up to larger detector counts. Like TESes, KIDs can be fabricated on silicon wafers using thin-film lithography. However, each KID pixel, which is a superconducting resonator, has a unique frequency and a narrow bandwidth. This is an advantage of KIDs over TESes because large numbers of KIDs can be read out on a single transmission line using microwave frequency-division multiplexing (mFDM) without requiring complex SQUID-based readout and assembly.

In the place of SQUIDs, KID readout systems typically use fast FPGA-based\cite{Bourrion_2016, vanRantwijk_2016, Golwala_2012, Gordon_2016} or GPU-based\cite{usrpgpu, usrpgpugithub} digital electronic systems to generate probe tones that are sent down a coaxial line with filters and attenuators to excite the KID array. The detector responses are then amplified using a cryogenic Low Noise Amplifier (LNA) before passing out of the cryostat where they are demodulated and digitized into detector timestreams\cite{Mauskopf_2018}. 

Thermal Kinetic Inductance Detectors (TKIDs) are a variation of Kinetic Inductance Detectors and as such, they offer the same multiplexing benefits. A TKID integrates a superconducting resonator into a bolometer. Rather than tracking the changes in the resistance of the superconductor as TES bolometers do, the temperature variations of the thermally isolated island are measured by the changes in the kinetic inductance of the superconducting resonator circuit. Because all the power is thermalized on the bolometer island, a TKID's absorber can be electrically decoupled from the resonator circuit, unlike in most KID designs. This gives TKIDs an additional degree of engineering flexibility since the resonator and the absorber can be optimized independently. 

TKIDs have been an active area of research since Sauvageau's\cite{sauvageau1989superconducting} early work on kinetic inductance thermometry. TKIDs have been developed as energy detectors for X-ray imaging spectroscopy\cite{tkidXray} to simultaneously achieve high spatial and energy resolution and for thermal X-ray photon detection\cite{quaranta2013x}. TKIDs have also been used as THz radiation detectors\cite{arndt2017optimization, timofeev2013submillimeter, dabironezare2018dual, wernis2013characterizing} operating at kelvin-range temperatures. However, TKIDs have not yet been demonstrated as power detectors for millimeter-wave instruments that operate at sub-kelvin temperatures.

The measurement of the polarization anisotropies of the Cosmic Microwave Background (CMB) is one of the most important use cases for bolometric power detectors working in the millimeter-wave regime. CMB observations have strict requirements on detector sensitivity, stability, and scalability, especially for future experiments targeting high detector counts \cite{cmbexperimentconsiderations, cmbs4}.

This paper therefore describes the performance of dark prototype devices that demonstrate the feasibility of using TKIDs for CMB observations. Our TKID design targets the detector performance requirements needed for CMB observations at 150 GHz. The devices discussed in this paper are an improvement on our initial design\cite{Steinbach2018} and demonstrate noise performance appropriate for background-limited ground-based CMB measurements at 150 GHz.

\section{\label{sec:devicetheory} Device Design and Expected Performance}
Like all bolometers, TKIDs consists of a thermally isolated island on which absorbed optical power $P_{\mathrm{opt}}$ is converted to heat. The absorbed power causes the island temperature $T$ to rise above the temperature of the heat sink held at a temperature $T_{\mathrm{bath}}$. The thermal island has a heat capacity $C(T)$ and is attached to the heat sink via a weak link with a thermal conductance $G(T)$. The temperature gradient across the thermal link creates a net power flow $P_{\mathrm{leg}}$ through the weak thermal link from the island to the heat sink. 

The readout circuit of a TKID consists of a parallel LC resonator built up of lumped element components. A large capacitor $C$ sets the resonance frequency $f_r$, and smaller additional capacitors $C_c$ couple the resonator to both the microstrip transmission line and to ground. The inductor, which acts as the thermometer, is on the island. The total inductance $L$ has a geometric inductance $L_g$ due to the shape of the inductor wire and a kinetic inductance $L_k$ from its superconductivity. 

The kinetic inductance of a superconducting film with thickness $t$, length $l$ and width $w$, for a total volume $V_{sc} = l \cdot w \cdot t$ can be determined from the normal sheet resistance $R_s$ of the film and its superconductor gap energy $\Delta$ using the relation $L_k = \left(l/w\right) \cdot L_s$ where $L_s = \hbar R_s/\pi\Delta$. This equation also defines the surface kinetic inductance $L_s$ that is often specified when doing electromagnetic simulations of superconducting resonators. The gap energy is in turn determined by the superconducting transition temperature $T_c$ via $\Delta\ =\ 1.763\ k_B T_c$. Changes in $L_k$ with temperature shift  both the resonance frequency and the internal quality factor $Q_i$ of the resonator. A large kinetic inductance fraction, $\alpha_k = L_k/(L_g + L_k)$ gives a large responsivity to thermal fluctuations\cite{jonasreference}. This is achievable by either minimizing the geometric inductance by design, or alternatively by choosing a material with a large resistivity or a large $T_c$.

\begin{figure*}[!htbp]
    \centering
    \includegraphics[width=0.98\textwidth]{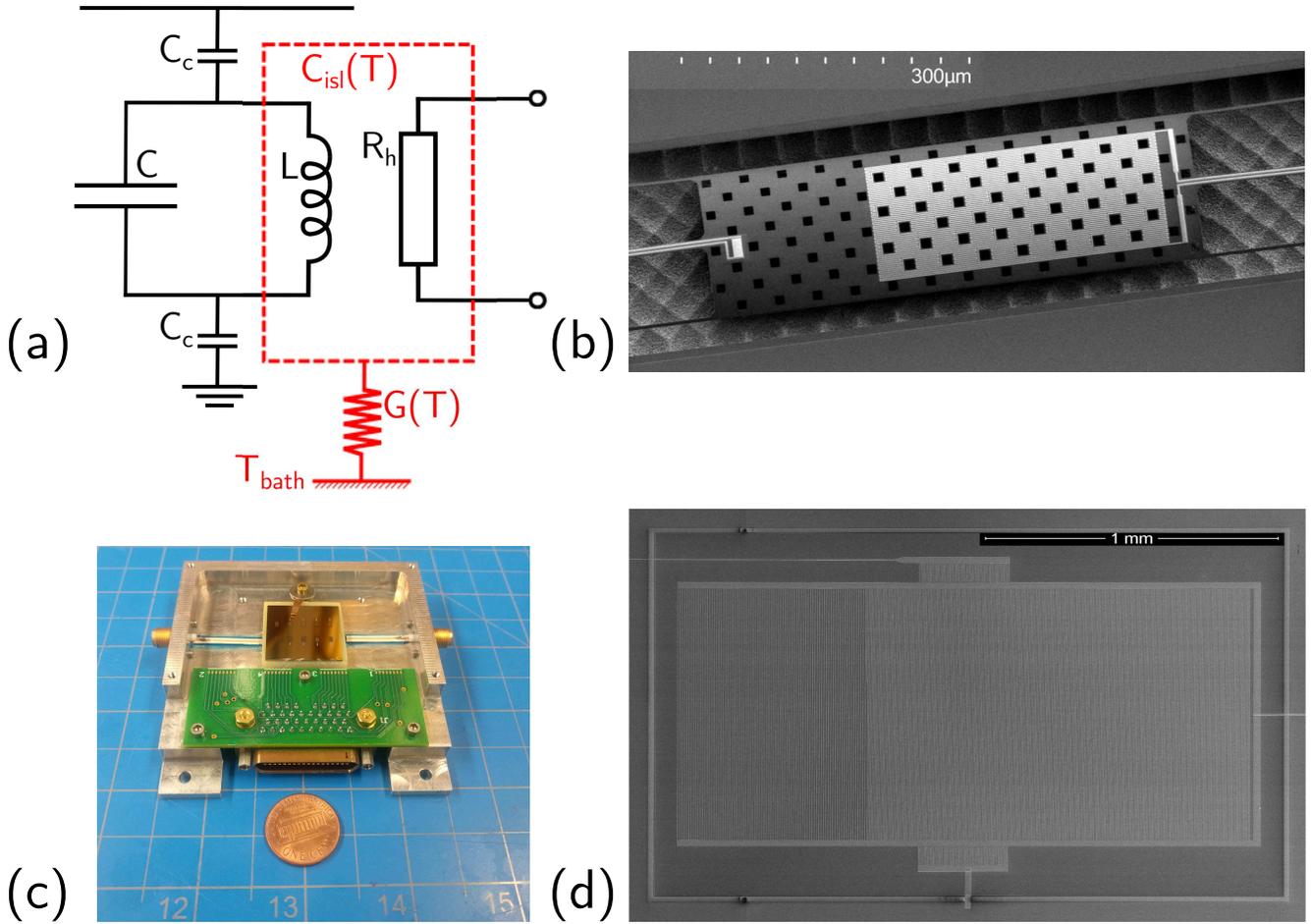}
    \caption{(a) Schematic of a TKID bolometer. The thermal circuit is shown in red while the electrical system is depicted in black. $R_h$ represents an Au resistor where we can deposit electrical power to simulate optical loading and to directly measure the power-to-frequency responsivity.\label{fig:tkidschematic} (b) SEM micrograph of a released bolometer showing the inductor on the right and the Au heater on the left. The square holes on the island are used in the Xe$\mathrm{F}_2$ release process. (c) Photograph of a single TKID prototype chip with the bias circuit.(d) SEM micrograph of the TKID capacitor and coupling capacitors with a section of the feedline visible.}
    \label{fig:allschematics}
\end{figure*}

The bolometer also includes an absorber element on the island that receives incident radiative (optical) power. In our dark device design, we use a Au resistor $R_h$ which can be biased through an external circuit to simulate optical loading on the island. The absorber is in thermal contact with but electrically isolated from the inductor. With this design, the absorber can be modified with no impact on the resonator. Because of this, a TKID can be used as a drop-in replacement for a TES in a bolometer in order to take advantage of already developed radiation-coupling technologies such as planar phased-array antennas, horns or lenslets\cite{rogerTES, polarbear, advancedACT}. Figure \ref{fig:tkidschematic}a shows a schematic of the thermal and electrical circuit of a TKID pixel as an aid to the discussion presented here.  

In the rest of this section, we build up the device model of a TKID bolometer in order to make predictions of their expected performance and noise properties. Our goal is to quantify the possible sources of noise that arise in addition to photon noise during observations. We expect detector noise contributions from 4 sources: phonon noise, generation-recombination (gr) noise, amplifier noise and two-level system (TLS) noise. Each of these noise terms is well covered in existing literature \cite{jonasreference, mather1, mauskopfreview} and we will therefore focus mainly on the conditions under which the total detector noise is dominated by the phonon noise which is still smaller than the expected photon noise level. Lindeman\cite{lindeman2014resonator} also gives a discussion of the noise sources in a resonator bolometer under a bolometer matrix formalism.

\subsection{\label{subsec:thermal} Thermal Architecture}
Bolometer theory is well established\cite{mather1, mather2, richards1} and we apply it to inform our predictions of TKID performance. There are a few notable distinctions between the operation of TKIDs and the more familiar TES bolometers that we will draw attention to.

Ignoring the noise terms at present, the thermal response of a TKID bolometer is governed by the differential equation:

\begin{equation}\label{eqn:thermodiff}
    C \frac{d T}{d t} =  - P_{\mathrm{leg}} + P_{\mathrm{read}} + P_{\mathrm{opt}},
\end{equation}
where $P_{\mathrm{read}}$ is the readout power dissipated by the inductor. Although the inductor is superconducting, it has a non-zero AC resistance. The net heat flow through the thermal link is modelled as $P_{\mathrm{leg}} = K \left(T^{n} - T_{\mathrm{bath}}^{n}\right)$, with $K$ a coefficient, and $n$ the power-law index. The thermal conductance of the thermal link is the derivative of $P_{\mathrm{leg}}$ with respect to the island temperature, $G(T) = n\ K\ T^{n-1}$. $P_{\mathrm{read}}$ is further discussed under section \ref{subsec:electrical} on the electrical properties of the bolometer.

Unlike TES bolometers, we operate TKIDs in the regime where $P_{\mathrm{read}} \ll P_{\mathrm{opt}}$, and therefore with negligible electro-thermal feedback. The equilibrium condition, $P_{\mathrm{opt}} = P_{\mathrm{leg}}$ sets the operating temperature $T_o$ at a given bath temperature. We target a 380 mK operating temperature from a 250 mK bath temperature, accessible with a $^3$He cooler. Importantly, the island temperature sets the quality factor of the resonators and in turn the number of resonators that can be packed in a given readout bandwidth. 

The CMB signal of interest can be described as small fluctuations in the optical power $\delta P_{\mathrm{opt}}$. We therefore expand equation \ref{eqn:thermodiff} to obtain the small signal response to small perturbations $\delta T$  about the operating temperature. The resulting equation in Fourier space is:

\begin{equation}\label{eqn:smallsignalthermal}
    \delta T(\nu) = \frac{\delta P_{\mathrm{opt}}}{G (T_o)} \cdot \frac{1}{1 + j 2 \pi \nu \tau_{\mathrm{bolo}}}.
\end{equation}
Here, $G(T_o)$ is the thermal conductance at operating temperature and is
determined by considering the total optical loading. The total optical loading
on the detector has contributions from both the microwave emission from the sky
and that of the telescope itself. The internal loading from the telescope is
usually minimized by design but is often still significant. As a reference, we
chose the representative measured optical loading, $P_{\mathrm{opt}} =\ 4.7\
\mathrm{pW} $ of the BICEP2 telescope which observed the CMB at 150 GHz from the
South Pole\cite{bicep2paper}. A 4.7 pW loading requires a thermal conductance $G (T_o) \sim 55$ pW/K. This conductance is realizable using silicon nitride films with a suitable geometry\cite{rogerTES}. Silicon nitride films can be used to achieve thermal conductances as low as the $10$ pW/K range to match a wide range of optical loading levels. Even lower thermal conductances down to $0.1$ pW/K can be achieved by incorporating phononic thermal filters in the thermal link design\cite{williams2018}. TKID bolometers can therefore be implemented under a wide range of potential loading conditions.

The bolometer time constant, $\tau_{\mathrm{bolo}} = C(T_o)/G(T_o)$, sets the bandwidth of the device. The device bandwidth, $\nu_{\mathrm{BW}} = 1/(2 \pi \tau_{\mathrm{bolo}})$ must be much larger than the science frequency band, i.e. $\nu_{\mathrm{BW}} > \dot{\theta}/\theta_{\mathrm{FWHM}}$ where $\dot{\theta}$ is the scan rate of the telescope and $\theta_{\mathrm{FWHM}}$ is the beam size. $\tau_{\mathrm{bolo}}$ is expected to be on the order of a few milliseconds, setting the device bandwidth at tens of Hz, which is fast enough for ground-based degree scale CMB observations. At fixed $G(T_o)$, the device bandwidth can be increased by modifying the island design to reduce the heat capacity.

We can now consider the thermal phonon noise which is generated by the random nature of the energy transport across the bolometer thermal link. The phonon noise level, expressed as a noise equivalent power is set by the thermal conductance $G(T_o)$ as:

\begin{equation}\label{eqn:phononNEP}
    \mathrm{NEP}_{\mathrm{ph}}^2 = 4 F(T_o, T_{\mathrm{bath}}) k_B T_o^2 G (T_o),
\end{equation}
where the factor $F(T_o, T_{\mathrm{bath}})$ accounts for the temperature gradient across the bolometer legs\cite{irwinhilton2005, mauskopfreview, mather1}. 

In comparison, the photon noise $ \mathrm{NEP}_{\mathrm{photon}}$ for a single-mode detector in the limit where the optical bandwidth $\Delta \nu_{\mathrm{opt}}$ is much smaller than the optical frequency $\nu_{\mathrm{opt}}$ is given by\cite{Zmuidzinas03}

\begin{equation}\label{eqn:photonNEP}
    \mathrm{NEP}_{\mathrm{photon}}^2 = 2 h \nu_{\mathrm{opt}} P_{\mathrm{opt}} + 2 \frac{P_{\mathrm{opt}}^2}{\Delta \nu_{\mathrm{opt}}}.
\end{equation}
To directly compare phonon noise to photon noise, we rewrite the phonon NEP by taking advantage of the fact that $P_{\mathrm{leg}} = P_{\mathrm{opt}}$ to eliminate $G(T_o)$ in favor of $P_{\mathrm{opt}}$. We find that $\mathrm{NEP}_{\mathrm{ph}}^2 = 4 \tilde{F}(T_o, T_{\mathrm{bath}}) k_B T_o P_{\mathrm{opt}}$, where $\tilde{F}(T_o, T_{\mathrm{bath}}) = F(T_o, T_{\mathrm{bath}}) \cdot n/(1 - \left(T_{\mathrm{bath}}/T_o\right)^{n})$. For our device parameters, $F(T_o, T_{\mathrm{bath}}) = 0.57$ and $\tilde{F}(T_o, T_{\mathrm{bath}}) = 2.4$, giving $\mathrm{NEP}_{\mathrm{ph}} = 16\ \mathrm{aW}/\sqrt{\mathrm{Hz}}$, which is much smaller than $\mathrm{NEP}_{\mathrm{photon}} = 45\ \mathrm{aW}/\sqrt{\mathrm{Hz}}$ at $\nu_{\mathrm{opt}} = 150$ GHz, $\Delta \nu_{\mathrm{opt}}/\nu_{\mathrm{opt}} = 0.25$. We stress that our device is designed so that phonon noise sets the detector noise limit at operating temperature rather than other noise sources such as quasiparticle generation-recombination noise as in KIDs\cite{jonasreference}.

\subsection{\label{subsec:electrical} Electrical Properties}
The temperature fluctuations of the bolometer island must be converted into electrical signals that can be readout. The small signal changes in the island temperature are read out by monitoring the forward transmission $S_{21}$ of the TKID resonator. At a readout frequency $f$ that is close to the resonance frequency $f_r$,\ $S_{21}(f)$ for a single-pole resonator is well described using the equation\cite{khalil, deng2013}:

\begin{equation}\label{eqn:s21}
    S_{21}(f) = 1 - \frac{Q_r}{Q_c}\frac{1}{1 + j 2 Q_r x},
\end{equation}
where $x = (f - f_r)/f_r$ is the fractional detuning of the probe signal away from the resonance frequency, $Q_r$ is the loaded resonator quality factor and $Q_c$ is the coupling quality factor. Given $ Q_c$ and $Q_i$,  $Q_r^{-1} = Q_i^{-1} + Q_c^{-1}$. $Q_i$ accounts for all loss mechanism intrinsic to the resonator while $Q_c$ is a measure of how strongly the resonator is coupled to the external circuit through the feedline and ground. Under optimal coupling, $Q_i = Q_c \implies Q_r = Q_i/2$. 

Changes in the island temperature $\delta T(t)$ induce fractional changes in the resonance frequency,  $\delta x = (f_r(t) - f_r)/f_r$ as well as in the resonator dissipation, $\delta_i = Q_i^{-1}(t) - Q_i^{-1}$. If these fluctuations are slower than the resonator bandwidth $f_r/2 Q_r$, we are in the adiabatic limit and the response of the resonator $\delta S_{21}(\nu)$ at Fourier frequency $\nu$ is given by

\begin{equation}\label{eqn:deltaS21}
    \delta S_{21}(\nu) = \frac{Q_i}{4} \chi_c \chi_g e^{-2 j \phi_g} \left[\delta_i(\nu) + j 2 \delta x(\nu) \right],
\end{equation}
where we have defined the coupling efficiency, $\chi_c \equiv {\textstyle \frac{4 Q_r^2}{Q_i Q_c}}$ such that $\chi_c\ =\ 1$ at optimal coupling. In addition, we also define the detuning efficiency $\chi_g \equiv \left[1 + 4 Q_r^2 x^2 \right]^{-1}$ with $\tan \phi_g = 2 Q_r x$.  Similarly, at zero detuning, $f\ =\ f_r \implies \phi_g\ =\ 0$ and $\chi_g\ =\ 1$. These quantities relate the probe signal power $P_g$ to the power dissipated by the resonator $P_{\mathrm{read}}$ as $P_{\mathrm{read}} = {\textstyle \frac{1}{2}} \cdot \chi_c \cdot \chi_g \cdot P_g$. At optimal coupling and zero detuning, half the input power is dissipated by the resonator\cite{jonasreference}. In the TKID case, only a portion of the resonator is on the thermal island, so the derived expression for $P_{\mathrm{read}}$ is an upper limit on the readout power contribution to the total loading on the island.

We can compute $\delta x$ and $\delta_i$ by considering the electrodynamics of Bardeen-Cooper-Schrieffe (BCS) superconductors\cite{tinkham}. Typically, we make the approximation that the superconductor state is fully specified once the quasiparticle density $n_{\mathrm{qp}}$ is known. The quasiparticle generation mechanism is a key distinction between standard KIDs and TKIDs. In KIDs, energy is injected into the quasiparticles directly via photons whereas in TKIDs, thermal phonons are the main quasiparticle generators. For aluminum TKIDs, the expected quasiparticle lifetime $\tau_{\mathrm{qp}}$ is in the tens to hundreds of microseconds range; much smaller than the thermal time constant. Since we take the thermal quasiparticle generation to be the dominant mechanism, the total quasiparticle density reduces to the thermal quasiparticle density $n_{\mathrm{th}}$ given by BCS theory as\cite{tinkham}:

\begin{equation}\label{eqn:thermalnth}
    n_{\mathrm{th}} = 2 N_0 \sqrt{2 \pi k_B T \Delta} \cdot \exp\left[-\frac{\Delta}{k_B T}\right].
\end{equation}

As expected, $n_{\mathrm{th}}$ depends on the temperature $T$, the gap energy $\Delta$ and the single-spin density of states at the Fermi level $N_0$. Changes in the quasiparticle density with temperature can be directly related to changes in the resonant frequency and dissipation using the Mattis-Bardeen equations by introducing $\beta$; the ratio of the frequency response to the dissipation response. $\beta$ is a function of the probe frequency $f$ and temperature $T$ and for a thermal quasiparticle distribution, it is given by the equation\cite{jonasreference}

\begin{equation}\label{eqn:beta}
    \beta(f, T) = \frac{1 + \sqrt{\frac{2 \Delta}{\pi k_B T}} \exp\left[- \frac{h f}{2 k_B T} \right] I_0\left[\frac{h f}{2 k_B T} \right]}{\frac{2}{\pi}\sqrt{\frac{2 \Delta}{\pi k_B T}} \sinh\left[\frac{h f}{2 k_B T} \right] K_0\left[\frac{h f}{2 k_B T} \right]}.
\end{equation}
$K_0$ and $I_0$ are the zeroth-order modified Bessel functions of the first and second order respectively. At our targeted $T_o\ =\ 380$mK and $f\ \sim\ 300$MHz, $\beta >> 1$. This means that the resonator frequency shift channel provides a larger signal for thermometry than the resonator dissipation. Using $\beta$, we find that $Q_i \delta_i = \left(\delta n_{\mathrm{qp}}/n_{\mathrm{qp}}\right)$ and $Q_i \delta x = \frac{1}{2} \beta (f, T)  \left(\delta n_{\mathrm{qp}}/n_{\mathrm{qp}}\right)$.

We are now in a position to combine the thermal and electrical responses to obtain the power-to-frequency responsivity $S$ of a TKID bolometer:

\begin{equation}\label{eqn:responsivity}
    S \equiv \frac{\delta f_r}{\delta P_{\mathrm{opt}}} = \frac{\partial f_r}{\partial x} \frac{\partial x}{\partial n_{\mathrm{qp}}} \frac{\partial n_{\mathrm{qp}}}{\partial T} \frac{\partial T}{\partial P_{\mathrm{opt}}} = f_r (T) \frac{\kappa(T) \beta(f, T)}{2\ Q_i\ G(T)\ T}
\end{equation}
where \[\kappa(T) \equiv \frac{\mathrm{d\ ln}\ n_{\mathrm{qp}}}{\mathrm{d\ ln}\ T} = \left(\frac{1}{2} + \frac{\Delta}{k_B T}\right).\]

\subsection{Other Noise Terms}\label{subsec:intrinsicnoiseterms}
The three remaining noise terms are: generation-recombination (gr), TLS, and amplifier noise from the readout chain. All the noise terms add in quadrature: $\mathrm{NEP}_{\mathrm{total}}^2 = \mathrm{NEP}_{\mathrm{photon}}^2 + \mathrm{NEP}_{\mathrm{ph}}^2 + \mathrm{NEP}_{\mathrm{gr}}^2 + \mathrm{NEP}_{\mathrm{amp}}^2 + \mathrm{NEP}_{\mathrm{TLS}}^2$. We consider each of these terms in turn and describe the conditions under which each of them remains sub-dominant to the phonon noise level.

\paragraph{Generation-recombination (gr) noise} \hspace{0pt} \\
Generation-recombination noise is present in TKIDs because quasiparticle production and recombination are random processes that fluctuate over time. Even so, the dynamics of the averaged quasiparticle population are well understood using non-equilibrium statistical mechanics\cite{grnoisepaper}. TKIDs operate at large quasiparticle densities where the quasiparticle recombination rate $\Gamma_{\mathrm{rec}}$ is quadratic in the quasiparticle density,

\begin{equation}\label{eqn:gammarec}
     \Gamma_{\mathrm{rec}} = \frac{1}{2} V_{\mathrm{sc}}\ R \ n_{\mathrm{qp}}^2,
\end{equation}
where $R$ is the recombination constant of the superconductor with volume $V_{\mathrm{sc}}$. For small perturbations in the quasiparticle density about its mean, quasiparticles have a characteristic lifetime $\tau_{\mathrm{qp}} = 1/ \left(R\cdot n_{\mathrm{qp}}\right)$. It is important to note that the energy relaxation time is half the quasiparticle lifetime\cite{Mauskopf_2018}. Accounting for this, the fluctuations in the quasiparticle number can be described by the one-sided power spectrum $S_{\mathrm{gr}}(\nu)$ given as\cite{grnoisepaper} 

\begin{equation}
    S_{\mathrm{gr}}(\nu) = \frac{2 \tau_{\mathrm{qp}}\ n_{\mathrm{qp}}\  V_{\mathrm{sc}} }{1 + (\pi \nu \tau_{\mathrm{qp}})^2}.
\end{equation}

The generation-recombination NEP  is the square root of the power spectrum divided by $V_{\mathrm{sc}}^2 \cdot \left(\partial n_{\mathrm{qp}} /\partial P_{\mathrm{opt}}\right)^2$,

\begin{equation}\label{eqn:grNEP}
    \mathrm{NEP}_{\mathrm{gr}} = \frac{G (T) T}{ n_{\mathrm{qp}}(T)\ \kappa (T)}\cdot\sqrt{\frac{2}{R V_{\mathrm{sc}}}}.
\end{equation}

Equation \ref{eqn:grNEP} shows that at high temperatures, gr noise is suppressed because the quasiparticle density increases exponentially with temperature. In addition, the responsivity, given in equation \ref{eqn:responsivity}, is independent of the superconductor volume. As a result, we are free to make the inductor volume large to further suppress the gr noise. This is an additional degree of flexibility for TKIDs, unlike many KID designs in which the inductor volume must be kept small to keep the optical responsivity high\cite{mccarrick, jonasreference}. The freedom to use a large inductor also allows us to use lower readout frequencies at a fixed capacitor size or alternatively, use smaller capacitors to achieve the same readout frequency.

\paragraph{Two-Level System noise} \hspace{0pt} \\
TLS noise in resonators is sourced by fluctuations in the permittivity of amorphous dielectric in the resonator\cite{glassesTLS}. These fluctuations couple to the electric field of the resonator. Unlike the noise terms already considered, there is no microscopic theory of TLS noise. Instead, we apply the semi-empirical TLS noise model which is extensively covered in Gao's thesis\cite{gaothesis}. TLS effects in the resonator also modify the internal quality factor and also introduce an additional temperature-dependent frequency shift. These two effects are given by

\begin{equation}\label{eqn:TLSQ}
    Q^{-1}_{\mathrm{TLS}} = F \delta_{\mathrm{TLS}} = F \delta_0\tanh \left(\frac{h f}{2 k_B T}\right)\cdot \frac{1}{\sqrt{1 + P_g/P_c}},
\end{equation}

\begin{equation}\label{eqn:TLSx}
\begin{split}
        x_{\mathrm{TLS}} &= \frac{F \delta_0}{\pi}\left[\operatorname{\mathbb{R}e}\left[\Psi \left(\frac{1}{2} + \frac{h f}{2 j \pi k_B T} \right)\right] - \ln \left(\frac{h f}{k_B T}\right)\right].
\end{split}
\end{equation}

Here, $\delta_{\mathrm{TLS}}$ is the TLS contribution to the dielectric loss tangent, $F$ is a filling factor that accounts for the fraction of the total electrical energy of the resonator that is stored in the TLS hosting material, $\delta_0$ is the loss tangent constant and $\Psi$ is the digamma function\cite{gaoTLS}. At a probe tone power $P_g$ and tone frequency $f$, TLS effects introduce a power dependence to the quality factor characterized by a critical power $P_c$. The fractional frequency shift is expected to be only weakly power dependent and positive with temperature increase. In the limit that $k_B T \ll h f$, the TLS loss drops off as $1/T$. Our target operating temperature is high enough to allow us to ignore the TLS effects on the resonator frequency and quality factor and only use the Mattis Bardeen relations. 

The TLS noise power spectrum $S_{\mathrm{TLS}}$ in $\mathrm{Hz}^{-1}$ in the limit of strong electric fields is given as\cite{gaothesis, jonasreference}

\begin{equation}
        S_{\mathrm{TLS}} \left[\mathrm{Hz}^{-1}\right] = \kappa_{\mathrm{TLS}} \left(\nu, f, T\right) \frac{\displaystyle \int \limits_{V_{\mathrm{TLS}}}\ |\vec{E}(\vec{r})|^3\  \mathrm{d}^3 r}{4 \left[\displaystyle \int \limits_{V} |\epsilon(\vec{r})\ \vec{E}(\vec{r})|^2\ \mathrm{d}^3 r\right]^2},
\end{equation}
where $\kappa_{\mathrm{TLS}} \left(\nu, f, T\right)$ captures the dependence on temperature and readout frequency, $\vec{E}(\vec{r})$ is the electric field, $\epsilon (\vec{r})$ is the dielectric constant, $V_{\mathrm{TLS}}$ is the volume of the TLS hosting media and $V$ is the total volume. The electric field terms exhibit the measured $P_g^{-1/2}$ dependence on the readout power. 

In order to compare TLS noise in devices with different geometry and operating conditions, it is more useful to use the microwave photon number $N$ instead of the electric field. $N = E/\left(h f_r\right)$, where the $E$ is the energy stored in the resonator. $E = \frac{1}{2} \cdot Q_i \cdot \chi_c \chi_g \cdot P_g/\left(2 \pi f_r\right)$, from the definition of the internal quality factor. We must also account for the known saturation of the power dependence of TLS effects at low electric fields\cite{gaoTLS}. We can include this saturation factor and make the temperature and readout frequency dependence of $\kappa_{\mathrm{TLS}}$ explicit by rewriting  $S_{\mathrm{TLS}}$ as

\begin{equation}
    S_{\mathrm{TLS}} \left[\mathrm{Hz}^{-1}\right] = \kappa_{\mathrm{TLS}, 0} \left(\frac{\nu}{300 \mathrm{MHz}}\right)^{-\alpha} \left(\frac{T}{380 \mathrm{mK}}\right)^{-\beta} \left(1 + N/N_c\right)^{-\gamma},
\end{equation}
where $\kappa_{\mathrm{TLS}, 0}$ is a constant that sets the overall TLS noise level. The exponents typically have measured values\cite{jonasreference} $\alpha = 1/2, \beta = 1.5-2\ \mathrm{ and }\ \gamma = 1/2$ although other values for the exponents $\alpha$ and $\beta$, have been suggested\cite{f1/3paper, interactingTLS, universalTLS, f1paper}. $N_c$ captures TLS saturation at $N \ll N_c$ with the correct limit when $N \gg N_c$. We estimate $N_c \sim 7 \times 10^6$ from measured TLS critical powers of our devices, $P_c \sim -95\ \mathrm{dBm}$. A few simple scaling relations between measured $\delta_{\mathrm{TLS}}$ and $\kappa_{\mathrm{TLS}, 0}$, which are useful for predicting TLS behavior, have been reported in literature\cite{universalTLS, f1paper, interactingTLS}. We do not refer to these but we instead consider measured noise levels for similarly designed devices presented in Figure 14 of Zmuidzinas' review\cite{jonasreference} to estimate $\kappa_{\mathrm{TLS}, 0} \sim 8\ \times\ 10^{-23}\ \mathrm{Hz}^{-1}$. We specify $T = T_o$ and $f = 300$MHz, to obtain an upper limit of $S_{\mathrm{TLS}}(1 \mathrm{Hz},\ T_o,\ -90 \mathrm{dBm}) \sim 4\ \times\ 10^{-19}\ \mathrm{Hz}^{-1} $.

We can obtain the TLS NEP by dividing the TLS power spectrum by the power-to-fractional-frequency-shift responsivity

\begin{equation}\label{eqn:tlsNEP}
    \mathrm{NEP}_{\mathrm{TLS}}^2 = \frac{1}{\left(\partial x/\partial P_{\mathrm{opt}}\right)^2} S_{\mathrm{TLS}}.
\end{equation}
In order to satisfy $\mathrm{NEP}_{\mathrm{TLS}}^2 \ll \mathrm{NEP}_{\mathrm{ph}}^2$, the following condition must hold:
\begin{equation}
    S_{\mathrm{TLS}} \ll 4 \chi k_B T_o^2 G(T_o) \left(\frac{\partial x}{\partial P_{\mathrm{opt}}}\right)^2.
\end{equation}

When we specify our design parameters, the condition reduces to $S_{\mathrm{TLS}} \ll 2.7 \times 10^{-17} \ \mathrm{Hz}^{-1}$. The condition is satisfied because of the high responsivity $\partial x/\partial P_{\mathrm{opt}} = 331\ \mathrm{ppm}/\mathrm{pW}^{-1}$ at $T_o$. The two orders of magnitude gap between the expected TLS noise level and the upper tolerable limit gives us confidence that TLS noise will have negligible impact on our devices during normal operation.

\paragraph{Amplifier Noise} \hspace{0pt} \\
The last noise contribution to consider is the additive noise of the  amplifier. For an amplifier with noise temperature $T_N$ and and at readout power $P_g$, the NEP contribution is given by\cite{jonasreference}

\begin{equation}
    \mathrm{NEP}_{\mathrm{amp}} = \frac{1}{\left(\partial x/\partial P_{\mathrm{opt}}\right)} \cdot \frac{Q_c}{2\ Q_r^2}\sqrt{\frac{k_B T_N}{P_g}}.
\end{equation}

The amplifier noise contribution can be made small by using an amplifier with a low enough noise temperature or by biasing the resonators with a large readout power. Cryogenic low noise amplifiers with $T_N < 10$ K are readily available commercially. However, we deliberately limit the bias power to operate the resonators in the linear kinetic inductance regime.

\begin{figure}
    \centering
    \includegraphics[width=0.48 \textwidth]{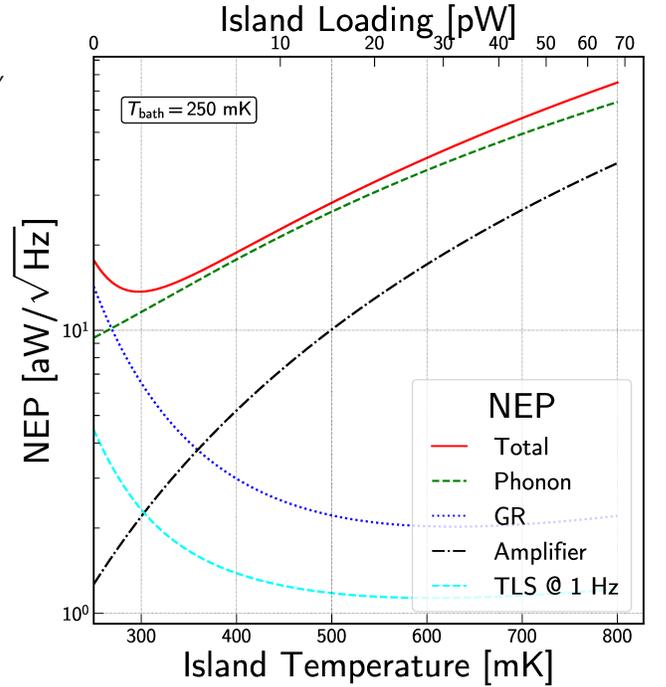}
    \caption{Detector noise model of an aluminum TKID as a function of the island temperature showing each noise term. TLS noise was modeled using $\kappa_{\mathrm{TLS}, 0} = 8\ \times\ 10^{-23}\ \mathrm{Hz}^{-1}$ and the amplifier noise was calculated at a 5 K noise temperature and -90dBm readout power. The bolometer properties were chosen to match those of the 337 MHz resonator in table \ref{tab:summarytable} and the resonator was taken to be optimally coupled at 380 mK. }
    \label{fig:NEPvstemperature}
\end{figure}

An optimized TKID bolometer has detector noise contributions, $\mathrm{NEP}_{\mathrm{gr}},\ \mathrm{NEP}_{\mathrm{TLS}}\ \mathrm{and}\ \mathrm{NEP}_{\mathrm{amp}}$ below the phonon noise. Figure \ref{fig:NEPvstemperature} shows the noise model of an aluminum TKID as a function of the island temperature. The figure shows the limits under which generation-recombination vs. phonon noise dominates in the resonator. The bolometer parameters were chosen to be suitable for 150 GHz ground-based observations of the CMB. The resonator was taken to be optimally coupled at 380 mK and the readout power was set at -90 dBm with a 5 K amplifier noise temperature.  

\subsection{Device Design}\label{subsec:design}
We can apply the results from the previous subsections to determine an optimal design for TKID bolometers. First, since the phonon noise term dominates the internal noise, then the internal noise at the operating temperature is only a weak function of $T_c$. This means that a wide range of materials with $T_c$ in the range $0.8 - 2$ K can be used as background-limited detectors. We chose aluminum with $T_c \sim 1.2 $K as our superconductor for ease of fabrication. Using higher $T_c$ materials could offer a multiplexing advantage because $Q_r$ at the operating temperature increases with $T_c$.

We designed devices with relatively low resonance frequencies around 300 MHz. We designed each test chip with 5 TKIDs with a 15 MHz frequency spacing between resonators. The resonator circuits are built out of lithographed, lumped element inductors and capacitors. A fixed inductor geometry was used for all the devices and each resonator has a unique main capacitor that sets the resonant frequency. Smaller coupling capacitors are used to set the coupling of the resonator to the readout line. Niobium with $T_c \sim 9$ K was used for all the capacitors and feedline structures so that the thermal response is solely attributable to the aluminum inductor. Figure \ref{fig:allschematics} shows a simplified schematic of a TKID as well as a scanning electron microscope image of the bolometer island with the inductor and heater in view. 

\section{Detector Fabrication}\label{subsec:fabrication}
We fabricated the detectors at the Microdevices Laboratory at the Jet Propulsion Laboratory (JPL) on 500 $\mu$m thick, high resistivity Si wafers\cite{rogerTES}. A low stress silicon nitride (LSN) layer, a niobium ground plane (GP) and a $\mathrm{SiO}_2$ inter-layer dielectric (ILD) layer are deposited over the silicon. The LSN layer is then patterned to form the thermal island which is released using a Xe$\mathrm{F}_2$ etch process. The island is mechanically anchored to the main wafer via six LSN legs each about 10 $\mu$m wide. These legs make the weak thermal link from island to wafer. By selecting the length of the bolometer legs, we can also tune the thermal conductance. On each test chip, one device was fabricated with no island and the other 4 with bolometer leg lengths 100 $\mu$m, 200 $\mu$m, 300 $\mu$m and 400 $\mu$m. The island size was fixed to $\sim$100 $\mu$m by $\sim$500 $\mu$m for all the devices. This gives an expected heat capacity of 0.06 pJ/K, after accounting for both the dielectric (LSN and ILD) and metal (Nb and Al) layers\cite{heatcapacitySiN, heatcapacitySiO2, heatcapacityAlandNb}. The heat capacity is set by island volume of about 30,000 $\mu$m$^3$, which is much larger than the combined volume of the bolometer legs of about 5,000 $\mu$m$^3$ for the longest leg bolometer. As a result, we expect that all the devices would have the same the heat capacity. A small gold resistor added on the island is used for calibration and noise measurements(see Figure \ref{fig:allschematics} (b)).

The aluminum layer of the inductor is deposited and patterned on the LSN . The meandered inductor traces are 50 nm thick and $1\ \mu$m wide with 1 $\mu$m spacing between the lines for a total volume of 810 $\mu \mathrm{m}^3$.  Sonnet\cite{sonnet} simulations done assuming a surface inductance $L_s = 0.27\ \mathrm{pH}/\square$ , give a predicted geometric inductance, $L_g$ of about 5 nH and a kinetic inductance fraction, $\alpha_k = 0.42$. The low film resistivity of aluminum limits the achievable kinetic inductance fraction. 

The inter-digitated capacitors (IDC) are deposited directly on the crystalline Silicon wafer. The bare silicon is exposed by etching a large via through the ground plane (GP) and LSN layers. We did this to reduce the presence of amorphous dielectric that hosts two-level system\cite{gaoTLS} (TLS) effects from the capacitors. The via is large to minimize stray capacitive coupling between the ground plane and the edges of the capacitor, as determined by Sonnet simulations. An etch-back process was used to pattern the capacitors because liftoff often leaves flags that can potentially host TLS states.

\section{Experimental Setup}
We tested the devices in a Model 103 Rainier Adiabatic Demagnetization Refrigerator (ADR) cryostat from High Precision Devices (HPD)\cite{hpd} with a Cryomech PT407 Pulse Tube Cryocooler\cite{cryomech}. We used an external linear stepper drive motor to run the Pulse Tube in order to avoid RF pulses from a switching power supply. For mechanical stiffness, the ultra cold (UC) and intra cold (IC) stages of the cryostat are supported using long diameter Vespel tubes between the 4K stage and the IC and stiff titanium 15-3 alloy X-shaped crossbars between the IC and UC stages. Copper heat straps make the thermal contacts between the stages and the cold heads. The ADR reached a base temperature of 80 mK on the UC stage when the test chip was installed.

The RF connections between the UC and IC stages are through niobium-titanium (NbTi) coaxial cables with 10dB, 20dB and 30dB attenuators installed at the UC, IC and 4K stages respectively on the transmit side. At each thermal stage, the coaxial connections are heat sunk to the stage. A cold Low Noise Amplifier (LNA) is mounted at the 4K stage. The cold LNA is a SiGe HBT cryogenic amplifier from Caltech (CITLF2)\cite{lnaref} with a measured noise temperature of 5.2 K with a 1.5 V bias. We also use a second amplifier with a noise temperature of 35 K at 5.0 V bias at room temperature. 

For our data acquisition, we used the JPL-designed GPU accelerated system  built on the Ettus Research USRP software defined radio (SDR) platform \cite{usrpgpu, usrpgpugithub}. 10 Gbit Ethernet connects the SDR to an Nvidia GPU, which handles the computationally heavy demodulation and analysis tasks in place of an FPGA. The SDR platform uses a 14 bit ADC and a 16 bit DAC and provides 120 MHz RF bandwidth.

The chip holder was sealed up with aluminum tape to make it light tight\cite{radiationloss}. In addition, we anchored a Nb plate onto the UC stage to provide some shielding of the chip from the magnetic field of the ADR magnet.

\section{\label{sec:results} Results}
For the device discussed here, we achieved a yield of 4/5 resonators. All the resonances were found in the frequency range 300-360 MHz and with close to the 15 MHz design spacing. The heaters of the 300, 200 and 100 $\mu$m leg bolometers were wired up to make calibrated noise and responsivity measurements.

\subsection{Film Properties}
Using four-point measurements of a test structure on chip, we measured $T_c = 1.284 \ K$, implying a superconducting gap energy, $\Delta = 1.95 \times 10^{-4}\ \mathrm{eV}$.  The measured sheet resistance was $R_s = 0.25\ \Omega/\square$. $R_s$ and $\Delta$ together give a surface inductance, $L_s = 0.27\  \mathrm{pH}/\square$. Taking this value of $L_s$ and accounting for the geometric inductance from simulations, gives an expected kinetic inductance fraction, $\alpha_k = 0.42$.

\subsection{Resonator Properties and Modeling}
We initially characterized the resonators using readout power sweeps at a fixed bath temperature of 80 mK and varying the power in the range, -110 dBm to -80 dBm. A key goal of this measurement was to determine the power level at the onset of the kinetic inductance non-linearity. In addition, changes in the quality factor with readout power probe two-level system effects. 

We fit $S_{21}$ using Swenson's nonlinear resonator model\cite{swenson}. However, we found that our resonators could not be described adequately using the ideal $S_{21}$ presented in equation \ref{eqn:s21}. In general, parasitic inductance and capacitance, wirebond inductances, line mismatch and other effects modify the resonator line shape. In our experimental setup, these effects show up as an asymmetry in the resonator. An asymmetric line shape can be accounted for by making the coupling quality factor complex ($Q_c \rightarrow \hat{Q}_c$)\cite{khalil, deng2013}; introducing an additional fitting parameter. $Q_r$ and the real and imaginary parts of $\hat{Q}_c$ are determined from the best fit to $S_{21}(f)$ and the internal quality factor $Q_i$ is then recovered using $Q_i = \left(Q_r^{-1} - \mathbb{R}e\left[\hat{Q}_c^{-1}\right]\right)^{-1}$. This relation motivates identifying 2 real quantities out of the complex $\hat{Q}_c$: the first being $Q_c = 1/\mathbb{R}e\left[\hat{Q}_c^{-1}\right]$ so that the equation $Q_r^{-1} = Q_i^{-1} + Q_c^{-1}$ holds; and the second, a coupling angle $\phi_c$ so that $\hat{Q}_c^{-1} = Q_c^{-1}\left(1\ +\ j\ \tan \phi_c\right)$.

The coupling angle $\phi_c$ captures the resonator asymmetry since as $\phi_c \rightarrow 0,\ \hat{Q}_c \rightarrow Q_c$. Our resonators have a large $\phi_c$ in the range $0.4 - 0.7$ radians. As tested by further simulations, we traced the asymmetry to about 7 pF of capacitive loading from short lengths of transmission line that branch from the readout line to the coupling capacitors. 

Even up to -80 dBm we found that the bifurcation parameter defined by Swenson $a < 0.8$, implying that the resonators were always below the threshold for the bifurcation\cite{swenson}. Informed by this, we chose a readout power of -90 dBm for all the other measurements that were done at a fixed readout power. At -90 dBm, the amplifier noise is sufficiently suppressed while keeping the readout power smaller than the heater loading on the TKID.

$Q_i$ varies as a function of both the readout power $P_g$ and the temperature $T$. We chose the following model to describe $Q_i$:
\begin{equation}\label{eqn:qivspvst}
    Q_i^{-1}(T, P_g) = Q_{\mathrm{TLS}}^{-1}(T, P_g)+ Q_{\mathrm{MB}}^{-1}(T) + Q_{i,0}^{-1} ,
\end{equation}
where $Q_{\mathrm{TLS}}^{-1}(T, P_g)$, is given in equation \ref{eqn:TLSQ} and $Q_{\mathrm{MB}}^{-1}(T)$ is the Mattis-Bardeen (MB) prediction\cite{jonasreference} and the third parameter $Q_{i,0}$ captures the saturation of $Q_i$ at high power. 

\begin{figure}[!htbp]
    \centering
    \includegraphics[width=0.48\textwidth]{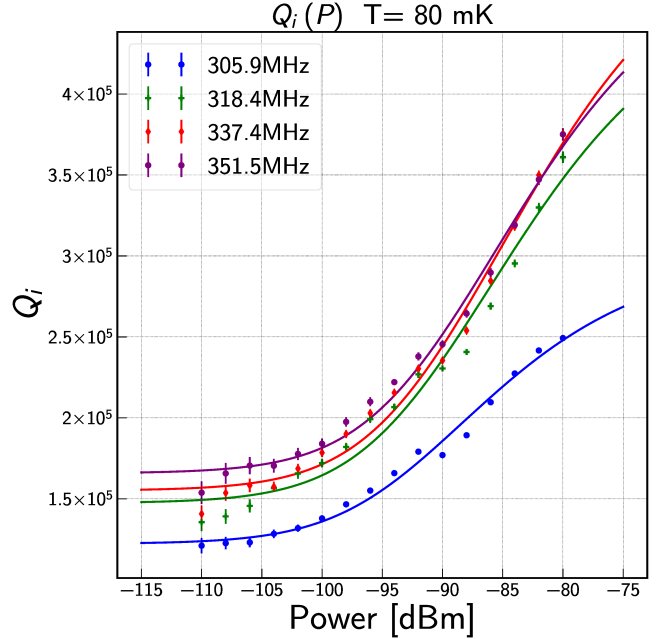}
    \caption{Internal quality factor $Q_i$ as a function of the readout power at 80 mK. The dots are the data and the solid lines are the fits to the data using the model described in equation \ref{eqn:qivspvst}. }
    \label{fig:qivsPTLS}
\end{figure}

At 80 mK, the MB term is negligible and can be ignored for the power sweep measurements. Figure \ref{fig:qivsPTLS} shows $Q_i$ as a function of the readout power for 4 yielded resonators. All four resonators have the product $F \delta_0 \approx 4.6 \times 10^{-5}$. Three of the resonators have $P_c \approx -95.7$ dBm and $Q_{i,0} \approx 4.4 \times 10^5$. The lower $Q_i$ measured for the 305.9 MHz resonator implied a lower $Q_{i,0}$. Consistent with the TLS prediction, we found the fractional frequency shift with power to be negligibly small, peaking at about 5 ppm. Even so, TLS effects by themselves do not fully  explain  the  power  dependence because  we found that $Q_{i,0}$ was  needed in order to obtain good fits even though it is not physically motivated. $Q_{i,0}$ may possibly come from a power dependence that is complicated by non-equilibrium quasiparticle distribution effects. This is still being actively investigated\cite{Goldie_2012, Visser_2014}.

\begin{figure}[!htbp]
    \centering
    \includegraphics[width=0.48 \textwidth]{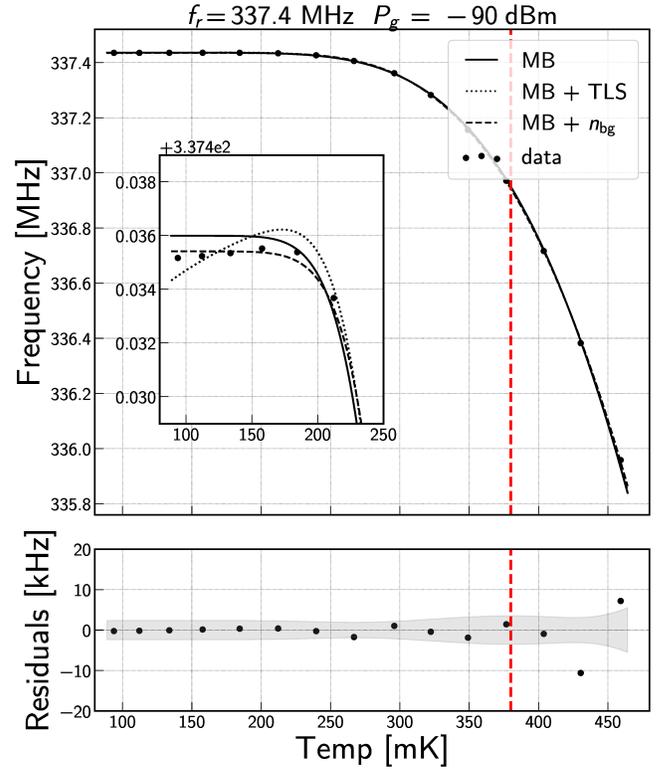}
    \caption{Fits of the frequency shift data for the 337.4 MHz resonator to three different models: only MB, MB + TLS and MB + a background quasiparticle density. The inset plot has a much smaller x-axis range to better show the differences in the fits at low temperatures. The lower plot gives the fit residuals (black dots) and the one-sigma error obtained from the covariance of the fit parameters for the MB + background quasiparticle model (gray). The red dashed line is our target operating temperature of 380 mK.}
    \label{fig:reso0xvsT}
\end{figure}

\begin{figure}[!htbp]
    \centering
    \includegraphics[width=0.48 \textwidth]{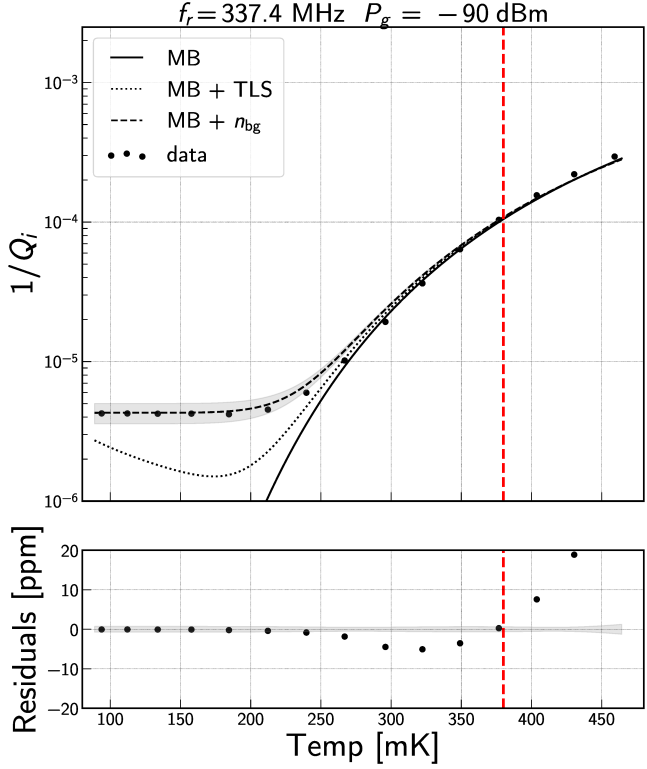}
    \caption{A comparison of the measured $Q_i$ to the best fit prediction for $Q_i^{-1}$ obtained from fitting the frequency shift data as shown in figure \ref{fig:reso0xvsT}. The best agreement between the data and the fit is with a MB + background quasiparticle density model. The lower plot gives the fit residuals (black dots) and the one-sigma error obtained from the covariance of the fit parameters for the MB + background quasiparticle model (gray). The red dashed line is our target operating temperature of 380 mK.}
    \label{fig:reso0qivsT}
\end{figure}

$S_{21}$ for each of the resonators was also measured as a function of the bath temperature. The readout power was fixed at -90 dBm and the temperature was swept up to 460 mK. This maximum temperature is high enough to break the degeneracy between the two MB parameters, $\alpha_k$ and $T_c$. At our operating temperature, these two parameters sufficiently describe our data as shown in figures \ref{fig:reso0xvsT} and \ref{fig:reso0qivsT} which summarize the fit results for the 337.4 MHz resonator.

At lower temperatures, however, the MB prediction fails because the quality factor levels off instead of increasing with a decrease in temperature. We compared three different models to describe this low temperature behavior: a pure MB prediction, a MB + TLS model and lastly, a model assuming a background quasiparticle population but no TLS. All three models were fit to only the frequency shift data and the best fit parameters were then used to make predictions of $Q_i$. The MB + TLS model best fit product $F \delta_0\ \approx 3.4\ \times 10^{-5}$ for all 4 resonators. However, this value of $F \delta_0$ gives a $Q_{\mathrm{TLS}}^{-1}$ that is too small for the data. In addition, if TLS effects were present, we would expect $Q_i^{-1}$ to increase with decreasing temperature below about 200 mK. Such an increase is not consistent with the data. 

A better fit was found using the background quasiparticle population $n_{\mathrm{bg}}$ model. In this case, total quasiparticle density is not simply the thermal quasiparticle density but rather it must be obtained by considering the balance between the quasiparticle generation and recombination rates\cite{grnoisepaper}. In addition, at low temperatures, empirical measurements show that quasiparticle lifetimes in Al resonators are described by the relation $\tau_{\mathrm{qp}} = \tau_{\mathrm{max}}/\left(1 + n_{\mathrm{qp}}/n_{\mathrm{qp}}^*\right)$. The two constants in the equation are $n_{\mathrm{qp}}^*$, which is the crossover density and $\tau_{\mathrm{max}}$, which is the observed maximum lifetime\cite{jonasreference, BarendsLifetime2008}. This new form of the quasiparticle lifetime has the expected inverse dependence on quasiparticle density in the limit of high $n_{\mathrm{qp}}$. Because of this, we can relate the two new constants to the recombination constant as  $R = 1/\left(\tau_{\mathrm{max}}\ n_{\mathrm{qp}}^*\right)$. The quasiparticle density is now modified to

\begin{equation}\label{eqn:nqpfull}
    n_{\mathrm{qp}}(T,\ n_{\mathrm{bg}}) = \sqrt{\left(n_{\mathrm{th}}(T) + n_{\mathrm{qp}}^*\right)^2 + n_{\mathrm{bg}}^2} - n_{\mathrm{qp}}^*
\end{equation}

Thermal and excess quasiparticles are equivalent in their effect on the electrodynamics of the superconductor\cite{gaoequiv}. As a result, we can describe combined the effect of having both by defining an effective temperature $T_{\mathrm{eff}}$ using the relation $n_{\mathrm{qp}}\left(T,\ n_{\mathrm{bg}}\right)\ =\ n_{\mathrm{th}}\left(T_{\mathrm{eff}}\right)$. The effective temperature was then used in place of $T$ for all temperature dependent terms that determine the frequency shift and loss with $n_{\mathrm{bg}}$ as an additional parameter in the fitting. We have no measurements of $n_{\mathrm{qp}}^*$ for these devices, but we fixed its value based on later quasiparticle lifetime measurements done on similar TKIDs for which we found $n_{\mathrm{qp}}^* \ =\ 518\ \mathrm{um}^{-3}$ and $R = 5.3\ \mathrm{um}^{3}\mathrm{s}^{-1}$. 

When comparing the best fit parameters from each of the three models, we found that the pure MB and the MB + TLS models give $T_c\ \approx\ 1.38$K and $\alpha_k\ \approx 0.55$, which are much higher values than we obtained from the film measurements. The background quasiparticle model gave $T_c\ \approx\ 1.32$K and $\alpha_k\ \approx 0.45$. On average over the 4 resonators, we found that $n_{\mathrm{bg}}\ \approx\ 700\ \mathrm{um}^{-3}$, corresponding to an effective temperature of about 150 mK. We estimate the recombination power, $P_{\mathrm{recomb}}$ from this population of quasiparticles as $P_{\mathrm{recomb}} = \frac{1}{2} \cdot \Delta \cdot R \cdot n_{\mathrm{qp}}^2 \cdot V_{\mathrm{sc}}\ =\ 9$ fW which is about 1\% of the readout power. The recombination power suggests that only a little additional loading on the bolometer is needed to account for the saturation of $Q_i$. A complete accounting of the low temperature quasiparticle dynamics in TKIDs as in many other KIDs has not been achieved. Even so, this does not affect TKID performance since our target operating temperature, $T_o = 380$ mK is well in the regime where our devices are fully characterized.

\begin{table*}[!htbp]
    \centering
    \begin{tabular}{c c c c c c c c c c  }
    $f_r$ [MHz] & $Q_c$ & $\phi_c$ [rad] & $T_c [\mathrm{K}]$ & $\alpha_k$ & $n_{\mathrm{bg}} [\mathrm{um}^{-3}]$ & $K$ [pW/$\mathrm{K}^{n}$] & n & $C_0$ [pJ/$\mathrm{K}^{\eta + 1}$] & $ \eta$ \\
    \hline \hline
    305.9 & 15164 $\pm$ 72 & 0.6455 $\pm$ 0.0035 & 1.33 $\pm$ 0.01 & 0.46 $\pm$ 0.01 & 708 $\pm$ 10 & 352 $\pm$ 2 & 2.962 $\pm$ 0.009 & 1.943 $\pm$ 0.213 & 1.920 $\pm$ 0.115 \\
    318.4 & 17675 $\pm$ 104 & 0.7601 $\pm$ 0.0034 & 1.32 $\pm$ 0.09 & 0.45 $\pm$ 0.01 & 711 $\pm$ 10 & 165 $\pm$ 1 & 2.754 $\pm$ 0.012 & 1.839 $\pm$ 0.325  & 1.945 $\pm$ 0.184 \\
    337.4 & 22556 $\pm$ 129 & 0.7526 $\pm$ 0.0037 & 1.32 $\pm$ 0.01 & 0.45 $\pm$ 0.01 & 627 $\pm$ 9 & 122 $\pm$ 1 & 2.862 $\pm$ 0.011  & 1.742 $\pm$ 0.380  & 1.914 $\pm$ 0.219 \\
    \hline
    \end{tabular}
    \caption{Summary of the measured parameters of 3 TKID bolometers. The error bars on $Q_c$ and $\phi_c$ were obtained from the spread in $Q_e$ over power sweeps. $\alpha_k, T_c,$ were obtained from the bath temperature sweep data fit with a MB + background quasiparticle model. The $K$ and $n$ values reported here are from measurements done at a 250 mK bath temperature. }
    \label{tab:summarytable}
\end{table*}

\subsection{Thermal Conductivity and Bolometer Time Constants}
By fixing the bath temperature and changing the power deposited by the heaters on the island, we can directly measure the thermal conductance and time constant.

The bath temperature, $T_{\mathrm{bath}}$ was fixed at 97 mK and $S_{21}$ was measured at each heater bias power $P$. The island temperature $T$ was then inferred from the resonance frequency shift using the best fit Mattis-Bardeen parameters $\alpha_k$ and $\Delta$ given in table \ref{tab:summarytable}, assuming that $T\ =\ T_{\mathrm{bath}}$ when the applied heater power is zero. The island temperature and bias power data are then fit to obtain the thermal conductivity coefficient $K$ and power law index $n$.

Consistent with other similar bolometers used in BICEP/Keck\cite{rogerTES}, we measured $n \sim 3$. An index $n < 4$ indicates that phonon system has reduced dimensionality. We can motivate this physically, under the assumption that ballistic phonon propagation dominates over scattering in thermal transport. In this limit, the thermal phonon wavelength is given by $\lambda_{\mathrm{ph}} = h c_s/k_B T$ where $c_s$ is the speed of sound\cite{rogerTES}. We expect $\lambda_{\mathrm{ph}}$ to vary from 3.5 $\mu$m at the 90 mK cold end to about 0.5 $\mu$m at the 450 mK hot end for a speed of sound, $c_s \sim 6500\ \mathrm{m  s}^{-1}$ in silicon nitride. The nitride layer is only 0.3 $\mu$m thick but the legs are 10 $\mu$m wide and 300 - 500 $\mu$m long. Consequently, we expect thermal behavior consistent with a 2 dimensional phonon gas. The longitudinal sound velocity in silicon dioxide is comparable to that of silicon nitride so the same argument holds for thermal conduction through the similarly thick ILD layer. With the different bolometer leg lengths on the chip, we verified that the coefficient $K$, scales inversely with the bolometer leg length as shown in Table \ref{tab:summarytable}. These observations are also supported by measurements of other silicon nitride bolometers that have reported similar values of $n$ for bolometers with leg lengths $<\ 400\ \mu$m and width $> 2\  \mu$m\cite{withington2017}. As an additional step, we repeated the thermal conductivity measurements at a second  bath temperature,~249 mK as shown in figure \ref{fig:calibPvsT270mK}. The extracted parameters were consistent with each other to about 10 \%. 

\begin{figure}
    \centering
    \includegraphics[width=0.48 \textwidth]{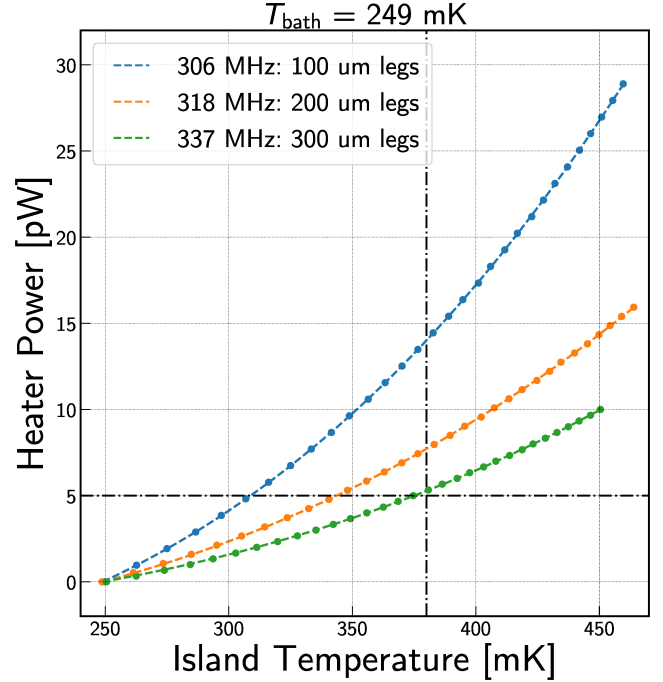}
    \caption{Heater power plotted against the island temperature for 3 TKIDs showing agreement between the data (points) and best fit model (lines). The data was taken at a 250 mK bath temperature. The difference in slopes between the 3 curves is due to the difference in thermal conductivity of the three bolometers. The horizontal and vertical lines are the target loading and operating temperature respectively. These are well matched by the 300um leg bolometer.}
    \label{fig:calibPvsT}
\end{figure}

\begin{figure}
    \centering
    \includegraphics[width=0.48 \textwidth]{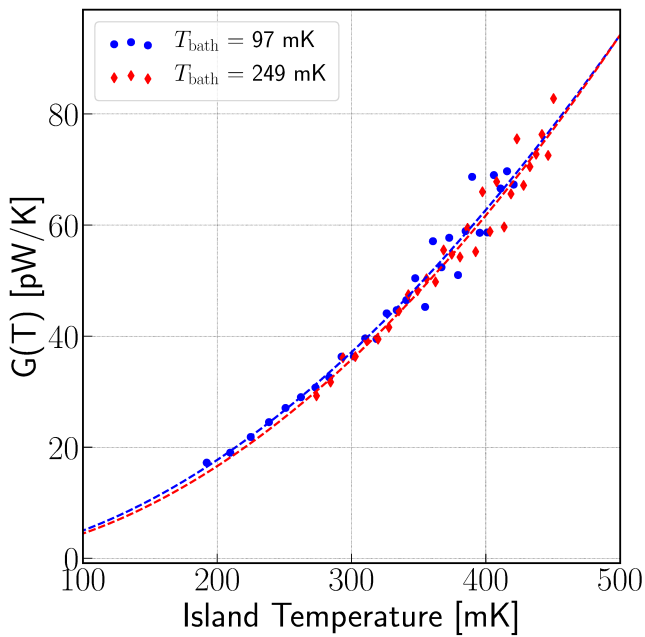}
    \caption{The thermal conductance of the 337 MHz bolometer extracted at 2 bath temperatures. The data are given as the filled blue circles and red diamonds while the dotted lines are the best fit to a power law model. The fits show that the parameter $K$ is independent of the bath temperature as expected.}
    \label{fig:calibPvsT270mK}
\end{figure}

We define the optimal power $P_{\mathrm{optimal}}$ as the power needed to achieve the target island operating temperature, $T_o = 380$ mK. $P_{\mathrm{optimal}}$ is dependent on the bath temperature and is chosen so that the saturation power equals the expected loading seen during actual observations. At a 250 mK bath temperature, the 300~$\mu$m bolometer with a resonant frequency of 337.4~MHz has $P_{\mathrm{optimal}}$ appropriate for the expected loading at the South Pole at 150~GHz as shown in figures \ref{fig:calibPvsT}. The 100~$\mu$m and 200~$\mu$m designs are suitable for operating at 270 GHz and 220 GHz respectively under the same sky conditions.

We measured the bolometer time constants using a DC bias voltage plus an additional sine wave excitation was applied across the heater. The amplitude of the sine wave was about 1 \% of the DC bias voltage. In addition, we synchronized the start of the data acquisition to the start of the sine wave excitation so that the phase of the input wave was always known. At each bias power, we stepped the excitation frequency of the input sine wave, $f_{\mathrm{exc}}$, from 1 Hz to 1000 Hz in 30 logarithmically spaced steps.

\begin{figure}
    \centering
    \includegraphics[width=0.48 \textwidth]{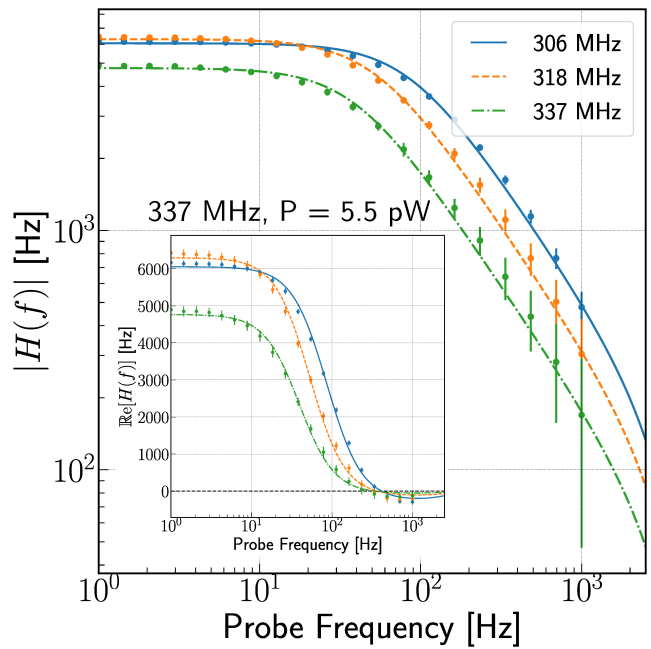}
    \caption{A comparison between the magnitude and the real part of the bolometer transfer function measured at an 85 mK bath temperature and heater power of 5.5 pW for the 337 MHz bolometer. The data points have error bars enlarged by a factor of 10. The solid lines show the best fit to the model in equation \ref{eqn:responsivity}. Inset is the real part of the bolometer response showing that response is modulated by the data filtering and additional time delays.}
    \label{fig:inphase}
\end{figure}

We convert from the raw I/Q timestreams to resonance frequency and quality factor timestreams using the measured $S_{21}$ parameters. By fitted the amplitude and phase of the change in resonance frequency with time, we obtained the complex bolometer transfer function, $H(f_{\mathrm{exc}})$ which we model as a single-pole low-pass filter with rolloff frequency $f_{3dB} = 1/2 \pi \tau_{\mathrm{bolo}}$. The measured transfer function also includes the effect of the anti-aliasing filter used to decimate the data as well as unknown time delays in the trigger signal from the USRP to the function generator. Figure \ref{fig:inphase}  show the magnitude of the bolometer response with the single pole rolloff.

\begin{figure}[!htbp]
    \centering
    \includegraphics[width=0.48\textwidth]{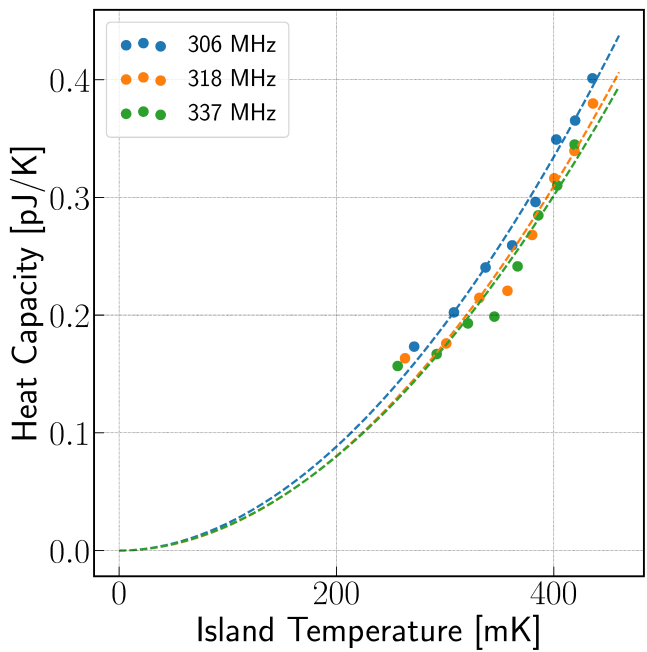}
    \caption{Heat capacity for 3 TKID bolometers as a function of the island temperature. All three bolometers have similar heat capacities despite having different leg lengths showing that the island volume is the dominant contribution to the total heat capacity of the bolometer. The dashed lines are power law fits to the heat capacity as a function of the island temperature. The best fit parameters from each bolometer are consistent with each other to within a 1 sigma uncertainty as reported in table \ref{tab:summarytable}.}
    \label{fig:heatcapacity}
\end{figure}

From these measurements, we conclude that the time constant for the 337 MHz TKID, $\tau \sim 4.5$ms is fast enough for ground-based degree scale CMB observations. Interestingly, the time constants are weakly dependent on the island temperature. We expect the heat capacity to follow a simple power law $C = C_0 T^{\eta}$ and given that $\tau = C/G$, then this implies that $C \sim T^2$ as summarized in table \ref{tab:summarytable}. Figure \ref{fig:heatcapacity} confirms that the bulk of the heat capacity is from the island itself and not the bolometer legs. Additionally, the thermal island is almost entirely dielectric so we expect $\eta = 3$. The measured heat capacity is about a factor of 3 larger than predicted (see section \ref{subsec:fabrication}). Even without electro-thermal feedback, TKIDs are still fast enough for our targeted science band. There are indications that our dry release process using XeF$_2$ contributes to the excess heat capacity and that using a wet release process could lead to faster bolometers for space-based applications\cite{beyerXEF2}.

\subsection{Responsivity}\label{subsec:responsivity}
Over large ranges of the optical power, the responsivity of a TKID bolometer as given in equation \ref{eqn:responsivity} is not constant. However, during normal CMB observations, the optical power is typically stable to a few percent. In this case, the responsivity can be approximated as being constant with a small non-linearity correction. The non-linearity level is similar to that of semiconducting NTD Ge bolometers that operate with small correction factors\cite{planckcalib1, planckcalib2}.

\subsection{Noise Performance}\label{subsec:noise}
The noise measurements were made by recording 10 minute long timestreams of the complex transmission at a 100~MHz native sample rate and then flat-averaging in 1000 sample blocks to a 100 kHz rate. A low noise, highly stable, Lakeshore 121 current source was used to DC bias the heater resistors. The power spectra of the resonance frequency timestream were estimated using the Welch method with a Hann window and a 50\% overlap. In order to convert from power spectra in frequency units to NEP, we directly measured the responsivity in short calibration noise acquisitions by modulating the DC bias with a 1\% 1~Hz square wave for 10 seconds. The responsivity is then estimated as the ratio of the change in frequency to the change in power dissipated. 

The data were taken at a 80 mK bath temperature because the UC stage was less stable at 250 mK. At the same island temperature, the phonon noise with a 250 mK bath temperature is 30\% higher than with a 80 mK bath. The measurement results presented here are focused on the 337 MHz resonator. 

\begin{figure}
    \centering
    \includegraphics[width=0.48 \textwidth]{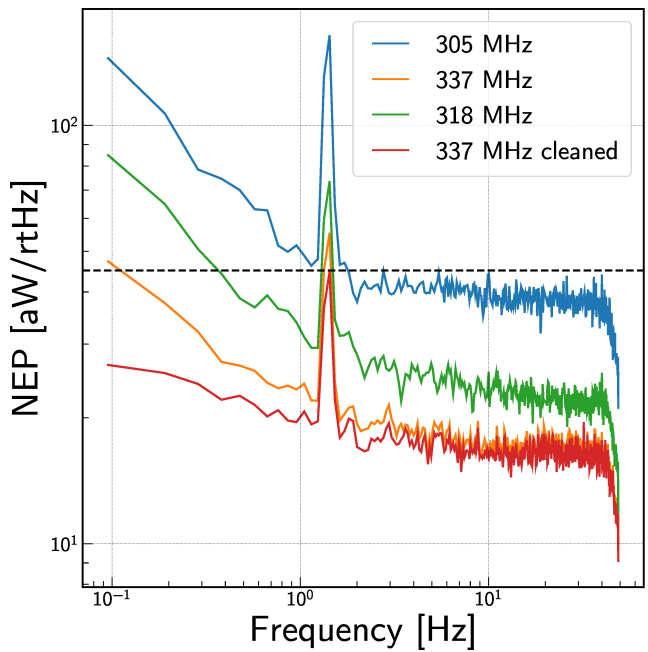}
    \caption{NEP spectra for 3 TKID resonators with 4pW loading on the 337 MHz resonator. The red line shows the spectrum of the 337 MHz resonator with common mode noise subtraction applied to suppress the noise at low frequencies. The large spike is the 1.4 Hz pulse tube line. The black dotted line shows the expected photon noise level for a single-mode detector at the South Pole observing in a band centered at 150 GHz with $\Delta \nu/\nu = 0.25$.}
    \label{fig:4pWcleanedNEP}
\end{figure}

To simulate pair differencing, which is usually done when making on-sky polarization measurements, we subtracted out all the noise in the 337 MHz resonator that was common-mode with the other resonators as shown in figure \ref{fig:4pWcleanedNEP}. The origin of the common-mode signal is still under investigation but is likely from either the thermal fluctuations of the stage or RFI susceptibility. From the filtered timestream we computed the NEP at a 4 pW heater bias level. Given the lower bath temperature, the island temperature is 320 mK.

\begin{figure}
    \centering
    \includegraphics[width=0.48 \textwidth]{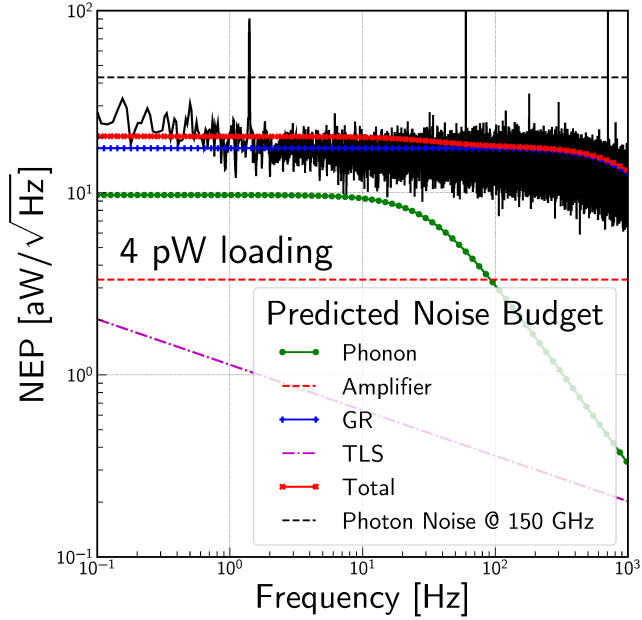}
    \caption{NEP of the 337 MHz resonator with a 4pW loading. Over-plotted are the estimated phonon, generation-recombination and amplifier noise contributions to the total noise based on the measured device parameters.}
    \label{fig:4pWnoisebreakdown}
\end{figure}

In the 4 pW noise dataset, the expected roll-off in the thermal responsivity at around 35 Hz is not visible. This indicates that at this loading, the device is the regime where the gr noise dominates, giving the single roll-off seen at around 1 kHz as shown in figure \ref{fig:4pWnoisebreakdown}. This figure also shows the comparison of the measured NEP to the noise model described in section \ref{sec:devicetheory}. However, our noise modeling is limited by our knowledge of the quasiparticle recombination constant $R$, which has not been directly measured for these devices. We instead modelled the noise by setting $R = 0.9\ \mu\mathrm{m}^3 \mathrm{s}^{-1}$ in order to match the gr noise roll-off seen at 4 pW loading with good agreement between the total noise from the model and the measured NEP.

\begin{figure}
    \centering
    \includegraphics[width=0.48 \textwidth]{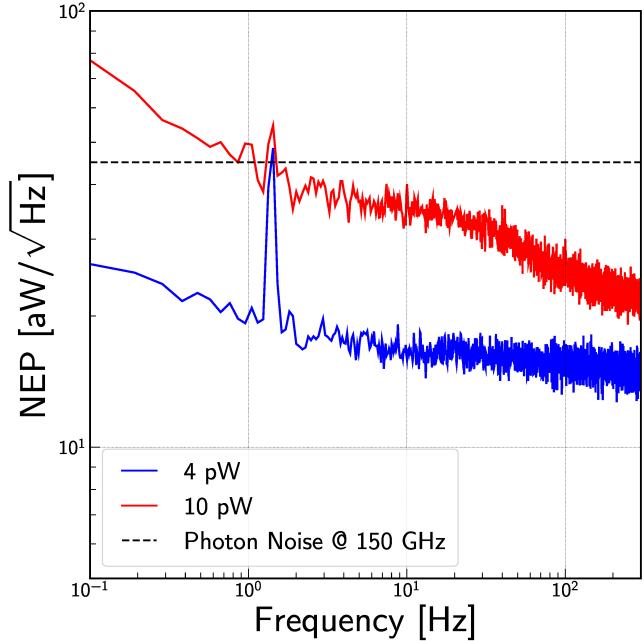}
    \caption{A comparison of the measured NEP for the 337 MHz resonator at the 4 and 10 pW loading levels after common mode noise subtraction. The large spike at 1.4 Hz is the pulse tube line. At 10 pW, there is a clear roll-off in the noise at around 35 Hz. This is consistent with the predicted thermal noise roll-off from the thermal conductivity and heat capacity measurements. At 4 pW, this roll-off is not visible suggesting that the gr noise is dominant detector noise term since the island temperature is much lower.}
    \label{fig:10pWnoisebreakdown}
\end{figure}

A second set of noise data was taken at a 10 pW heater bias that corresponds to an island temperature of 422 mK; much higher than the operating temperature. Figure \ref{fig:10pWnoisebreakdown} shows a comparison of the measured noise at the two loading levels after common-mode subtraction. Even at the elevated island temperature, the measured NEP is still below the expected photon NEP. However, the NEP measured at the 10 pW level is about 2 times greater than expected from the noise model. This discrepancy is still under investigation. Also of note is that at the 10 pW loading level we see a roll-off in the noise where the thermal response is expected to fall off.

At both loading levels, total detector noise is lower than the expected photon NEP of 45 aW/$\sqrt{\mathrm{Hz}}$ at the 4.7 pW loading for a ground-based CMB telescope observing in the 150 GHz band from the South Pole\cite{bicep2paper}. We conclude that the device noise is low enough that a  similar device coupled to an appropriate antenna\cite{chaolinTES, bicep_2015} would be background noise limited. Future work will focus on further understanding the gr noise and quasiparticle lifetimes in our devices.

\section{\label{sec:conclusions} Conclusions}
This paper presents a dark TKID design with detector noise that is low enough to produce background-limited performance when the device is appropriately coupled to an antenna tuned to observe in the 150 GHz frequency band from a ground-based CMB polarimeter. We've described the fundamental theory of TKID bolometer performance and compared it to actual devices. Our devices are fabricated using aluminum and niobium films deposited on a Silicon wafer using standard lithography techniques. We also presented the results from measurements of the film properties, bolometer characterization and dark noise measurements. Our device $Q_i$ and $f_r$ are well described using the Mattis-Bardeen theory at our operating temperature.

Future work will focus on further understanding quasiparticle lifetimes in our devices, increasing the number of devices and better understanding the common mode environmental noise in our measurement system. We will also work to better understand the performance of our devices under high cosmic ray backgrounds. As reported by several groups, arrays of KIDs can be susceptible to cosmic ray events\cite{cosmicrays1, cosmicrays2}. Cosmic rays passing through a wafer generate ballistic phonons that propagate across the wafer, affecting multiple detectors. Cosmic rays thus cause extended dead-times because a single event affects multiple devices and because one must wait across several time constants for the response to decay. Cosmic ray susceptibility becomes even more important in space applications where event rates are significantly higher\cite{planckcosmic1, planckcosmic2, spacecosmicrays1}. In TKIDs, the sensitive inductor is located on an island that is isolated from the entire wafer. The effective cross-section for interactions with cosmic rays is limited to the island instead of the entire wafer, greatly reducing our cosmic ray susceptibility. Other teams have reported similar results\cite{cosmicrays1}.

\section{\label{sec:acknowledgements} Acknowledgements}
The research was carried out at the Jet Propulsion Laboratory, California Institute of Technology, under contract with the National Aeronautics and Space Administration.  We thank the JPL Research and Technology Development  (RTD) program for supporting funds, as well as the Dominic Orr Graduate Fellowship in Physics at Caltech for supporting Mr. Wandui’s graduate research. We are grateful to Warren Holmes for his early guidance on shaping this project.

\section{\label{sec:dataavailability} Data Availability}
The data that support the findings of this study are available from the corresponding author upon reasonable request.

\section{References}
\nocite{*} 
\bibliography{bibliography} 

\end{document}